\documentclass{elsart}
\usepackage{epsfig,latexsym,amsmath,amsfonts}

\newcommand{\ENDPROOF}{$\hfill \Box$}
\newcommand{\ENDTHEOREM}{$\hfill \Diamond$}

\def\Gt{${\{\mathbb{G}(n,t)\}}_t \,$}
\def\VMAX{{V_l}^{max}}

\def\r{\right}
\def\l{\left}

\def\auteur#1{#1}
\onecolumn

\begin{document}
\runauthor{Baert, Ravelomanana and Thimonier}
\begin{frontmatter}
\title{On the growth of components with non fixed excesses}
\author[Amiens]{Anne-Elisabeth Baert},
\author[Lipn]{Vlady Ravelomanana} and
\author[Amiens]{Loys Thimonier}

\address[Amiens]{LaRIA CNRS EA 2083, Universit\'e de Picardie Jules-Verne \\
33, Rue du Moulin-Neuf \\
80000 Amiens, France}

\address[Lipn]{LIPN  CNRS UMR 7030, Universit\'e de Paris-Nord. \\
 99, av. J.B. Clement \\
 93430 Villetaneuse, France}

\date{\today}

\maketitle

\begin{abstract}
Denote by an $l$-component a connected graph with $l$ edges more than
vertices. We prove that the expected number of creations of $(l+1)$-component,
by means of adding a new edge to an $l$-component in a randomly growing
graph with $n$ vertices, tends to $1$ as $l,n$ tends to $\infty$
but with  $l = o(n^{1/4})$. We also show, under the same conditions on $l$ and
$n$, that the expected number of vertices that ever belong to an
$l$-component is $\sim (12l)^{1/3} n^{2/3}$.
\end{abstract}



\end{frontmatter}

\section{Introduction}
We consider here simple labelled graphs, i.e., graphs with labelled
vertices, without self-loops and multiple edges. A random graph is a pair
$(\mathcal{G}, P)$ where $\mathcal{G}$ is a family of graphs and $P$ is
a probability distribution over $\mathcal{G}$. 
Topics on random graphs provide a large and particularly active body of
research. For excellent books on these fields,
see \auteur{Bollobas} \cite{Bollobas} or \auteur{Janson} \cite{JLR2000}. 
In this paper, we consider the \textit{continuous time} random graph
model \Gt  which consists on assigning
a random variable, $T_e$, to each edge $e$ of the complete graph
$K_n$. The ${n \choose 2}$ variables $T_e$ are independent with
a common continuous distribution and  the edge set
of \Gt  is constructed with all edge $e$ such that $T_e \leq t$. Throughout this paper, a $(k,k+l)$ graph is one having $k$ vertices and $k+l$
edges. its \textit{excess} is $k$. A $(k,k+l)$ \textit{connected} graph is said an
\textit{$l$-component}. These terms
are due, as far as we know, respectively to \auteur{Janson} in \cite{Ja2000}
and to \auteur{Janson} \textit{et al}  in \cite{JKLP93}.

To obtain the results presented here, methods of the probabilistic model \Gt,
 studied in \cite{Ja2000}, are combined with asymptotic enumeration 
methods  (having the ``counting flavour'') developped 
by \auteur{Wright} in \cite{Wr80} and  by \auteur{Bender} \textit{et
 al.} in \cite{BCM90,BCM92}. The problems 
we consider are in essence combinatorial problems.
Often, combinatorics and probability theory are closely related and
the approaches given here furnish efficient uses of methods of asymptotic
analysis to get extreme characteristics of random labelled graphs.

Following \auteur{Janson} in 
\cite{Ja2000}, we define by $\alpha(l;k)$, the expected number of
times that a new edge is added to an $l$-component of order $k$ 
which becomes an $(l+1)$-component. This transition is denoted 
$l \rightarrow l+1$. It has been proved by Janson {\it et al.} 
in \cite{JKLP93} that with probability tending to $1$, there
is exactly one such transition, see Theorem 16 and its proof.
Our purpose in this paper is  to
give a different approach to obtain this result, when 
$l \to \infty$ with $n$. More precisely, we propose an alternative
method and direct calculations of the difficult results given
in \cite{JKLP93}, connecting these results to those of
Bender {\it et al.} in \cite{BCM90,BCM92}. To do this,
we evaluate $\alpha_l = \sum_{k=1}^{n}\alpha(l;k)$, 
the expected number of transitions $l \rightarrow l+1$, and show 
that, whenever $l \rightarrow \infty$ with $n$ but $l = o(n^{1/4})$,  
$\alpha_l \sim 1$. 

Moreover, let $V_l$ be the number of vertices that ever belong to
an $l$-component and $\VMAX$ the order of the largest $l$-component
that ever appears. We prove that, whenever $l\equiv l(n) \rightarrow \infty$ 
with $n$ but $l = o(n^{1/4})$, 
$V_l \sim (12l)^{1/3} n^{2/3}$ and 
$\VMAX = O(l^{1/3}n^{2/3})$. In particular,
these results improve the 
work of \auteur{Janson} and confirm his predictions
(cf. \cite[Remark 8]{Ja2000}).

\section{The Expected Number of Creations of $(l+1)$-excess Graphs}

When adding an edge in a randomly growing graph, there is a possibility
that it joins two vertices of the same component, increasing its excess.
Let $\alpha(l;k)$ be the expected number of times that a new edge is added
to an $l$-component of order $k$. Denote by $c(k,k+l)$ the number of
connected $(k,k+l)$-graph then we have the following lemma.

\begin{lem} \label{LEMMA1} For all $l \geq -1$ and $k \geq 1$, we have
\begin{equation}
\alpha(l;k) = {(n)}_k \frac{(k+l)!}{k!} c(k,k+l) %
 \frac{(k^2 -3k - 2l)}{2} %
\frac{(nk - k^2/2 - 3k/2 - l - 1)!}{(nk - k^2/2 - k/2)!}
\label{eq:LEMMA11}
\end{equation}
and for all $l$, $k$ and $n$ such that $l = O(k^{2/3})$, $1 \leq k \leq n$
\begin{eqnarray}
\alpha(l;k) = &  &  \frac{1}{2} \rho_l \frac{k^{(3l+1)/2}}{n^{l+1}} %
\exp\l(- \frac{k^3}{24 n^2} + \frac{l k^2}{8 n^2} + \frac{l k }{2 n} \r) %
                   \cr
 \times &  &\l( 1+ O\l(\frac{k}{n} + \frac{k^4}{n^3} + \frac{1}{k}\r) %
+ O\l(\frac{l^3}{k^2}+ \frac{l^{1/2}}{k^{1/2}}+ %
\frac{(l+1)^{1/16}}{k^{9/50}}\r) \r)
\label{eq:LEMMA12}
\end{eqnarray}
where 
\begin{equation}
\rho_l = \frac{1}{2} \sqrt{\frac{3}{\pi}}%
 \l(\frac{e}{12 l}\r)^{\frac{l}{2}} \l( 1 + O(1/l) \r) 
\label{eq:RHO-L}
\end{equation}
\ENDTHEOREM
\end{lem}

\noindent 
Before proving lemma \ref{LEMMA1}, let us recall the extension,
due to \auteur{Bender}, \auteur{Canfield} and \auteur{McKay} \cite{BCM92},
of \auteur{Wright}'s formula for $c(k,k+l)$, the asymptotic number of connected sparsely 
edged graphs \cite{Wr80}.

\begin{thm}[Bender-Canfield-McKay 1992] \label{THEOREM-BCM92}
There exists sequence $r_i$ of constants such that for each fixed
$\epsilon > 0$ and integer $ m > 1/\epsilon$, the number $c(k,k+l)$ 
of connected sparsely edged graphs satisfies
\begin{eqnarray}
c(k,k+l) = &  & \sqrt{\frac{3}{\pi}} \frac{w_l}{2} %
\l( \frac{e}{12l} \r)^{l/2} k^{k+(3l-1)/2} \cr
    \times &  & \exp \l(\sum_{i=1}^{m-2} \frac{r_i l^{i+1}}{k^i} %
+ O \l(\frac{l^m}{k^{m-1}} + \sqrt{\frac{l}{k}} + %
\frac{(l+1)^{1/16}}{k^{9/50}} \r) \r)
\label{eq:THEOREM-BCM92}
\end{eqnarray}
 uniformly for $l = O(k^{1-\epsilon})$. The first few values of
the constant $r_i$ are
\[
r_1 = - \frac{1}{2}, \,  r_2 = \frac{701}{2100},  \, r_3 = - \frac{263}{1050}, %
 \, r_4 = \frac{538 \, 859}{2 \, 695 \, 000} \, .
\]
\ENDTHEOREM
\end{thm}
Note that, the factor $w_l$ in (\ref{eq:THEOREM-BCM92}) is given, for 
$l  > 0$, by
\begin{equation}
w_l = \pi \frac{\Gamma(l)}{\Gamma(3l/2)} d_l \sqrt{\frac{8}{3}} %
\l( \frac{27l}{8e}\r)^{l/2} 
\label{eq:W_L}
\end{equation}
where $d_l = 1/(2 \pi) + O(1/l)$ and $w_0 = \pi/ \sqrt{6}$
(see \cite{BCM90,Wr80}). We also remark here that in lemma \ref{LEMMA1},
we restrict our attention to values of $l$ such that $l=O(k^{2/3})$ which
will be shown to be sufficient to obtain the result in theorem
\ref{THEOREM_EXPECTED_CREATION}. 

\noindent \textbf{Proof of lemma \ref{LEMMA1}.}
The proof given here are based on the works of \auteur{Janson} in
\cite{Ja93,Ja2000}. However, the main difference comes from the
fact that our parameter, representing the excess of the 
sparse components $l$, is no more fixed as in \cite{Ja2000}.
When a new edge is added to an $l$-component of order $k$,
there are ${n \choose k} c(k,k+l)$ manners to choose an $l$-component and 
${k \choose 2} - k - l$ ways to choose the new edge. Furthermore,
the probability that such possible component is one of
\Gt is $t^{k+l} (1-t)^{(n-k)k + {k \choose 2} - k - l}$ and with
the conditional probability $\frac{dt}{(1-t)}$ that a given edge is
added during the interval $(t,t+dt)$ and not earlier, integrating
over all times, we obtain
\begin{equation}
\alpha(l;k) = {n \choose k} c(k,k+l) \l(\frac{k^2 - 3k - 2l}{2} \r) %
\int_{0}^{1} t^{k+l} (1-t)^{(n-k)k + {k \choose 2} - k - l -1} dt
\label{eq:FORMULE_INTEGRAL}
\end{equation}
which evaluation leads to (\ref{eq:LEMMA11}).
For $1 \leq k \leq n$ and $l = O(k^{2/3})$, the value of the
integral in (\ref{eq:FORMULE_INTEGRAL}) is
\begin{equation}
\frac{(nk - k^2/2 - 3k/2 - l - 1)!}{(nk - k^2/2 - k/2)!} = %
k^{-k -l -1} (n - k/2)^{-k - l - 1} \l(1 + O(\frac{k}{n}) \r) \, .
\label{eq:LEMMA13}
\end{equation}
Furthermore,
\begin{equation}
\frac{{(n)}_k}{(n-k/2)^k} = \exp\l( -\frac{k^3}{24 n^2}\r) %
\l( 1 + O(k/n + k^4/n^3) \r)
\label{eq:LEMMA14}
\end{equation}
and obviously
\begin{equation}
{k \choose 2} - k - l = \frac{k^2}{2}\l( 1 + O(1/k)\r) \, .
\label{eq:LEMMA15}
\end{equation}
Thus, combining  (\ref{eq:LEMMA13}),(\ref{eq:LEMMA14})
and (\ref{eq:LEMMA15}) in (\ref{eq:LEMMA11}), we infer that
\begin{equation}
\alpha(l;k) = \frac{1}{2} \frac{(k+l)!}{k!} %
c(k,k+l) \frac{\exp \l( -\frac{k^3}{24 n^2} \r)}{(n-k/2)^{l+1} k^{k+l-1}}%
\l( 1+ O(1/k+k/n+k^4/n^3) \r) \, .
\label{eq:LEMMA16}
\end{equation}
Using Taylor expansions
\begin{equation}
\ln \l( \frac{(k+l)!}{k^l k!}\r) =  \frac{l^2}{2 k} + O( l^3/k^2 ) + O(l/k)
\label{eq:LEMMA17}
\end{equation}
which is sufficient for our present purpose $( l = O(k^{2/3}) )$.
Also, we get 
\begin{equation}
\frac{(n-k/2)^{l+1}}{n^{l+1}} = \exp \l( -\frac{lk}{2n} -
\frac{lk^2}{8n^2}\r) %
\l( 1+ O(k/n + l k^3/n^3) \r) \, .
\label{eq:LEMMA18}
\end{equation}
Note that $\frac{k^4}{n^3}$ dominates the term $ \frac{l k^3}{n^3}$ in 
(\ref{eq:LEMMA18}). Then, using the asymptotic
formula for the number $c(k,k+l)$ given by theorem \ref{THEOREM-BCM92} in
(\ref{eq:LEMMA16}), we obtain (\ref{eq:LEMMA12}).
\ENDPROOF

The form given by equation (\ref{eq:LEMMA12}), in lemma \ref{LEMMA1},
suggests us to consider the asymptotic behaviour of 
\[
\sum_{k=1}^{n} k^a %
\exp \l(- \frac{k^3}{24 n^2} + \frac{l k^2}{8 n^2} + \frac{l k }{2 n} \r) 
\]
where $a= \frac{3l+1}{2}$, $l \equiv l(n)$ as $n \rightarrow \infty$.

\begin{lem} \label{LEMMA2} 
As $n \rightarrow \infty$, $l \equiv l(n) = o(n^{1/4})$ 
and $a = \frac{3l+1}{2}$,  we have
\begin{equation}
\sum_{k=1}^{n} k^a %
\exp \l(- \frac{k^3}{24 n^2} + \frac{l k^2}{8 n^2} + \frac{l k }{2 n} \r)
\sim %
2^{a+1} 3^{(a-2)/3}  \Gamma \l(\frac{a+1}{3} \r) n^{2(a+1)/3} \, .
\label{eq:LEMMA2}
\end{equation}
\ENDTHEOREM
\end{lem}

\noindent \textbf{Proof.}  We start estimating the summation by an integral using, for e.g., the classical \auteur{Euler-Maclaurin} method for asymptotics
estimates of summations (see \cite{De Bruijn}),
\begin{equation}
\sum_{k=1}^{n} k^a exp \l(- \frac{k^3}{24 n^2} + \frac{l k^2}{8 n^2} + %
\frac{l k }{2 n} \r) \sim %
\int_{0}^{n} t^a \exp \l(- \frac{t^3}{24 n^2} + \frac{l t^2}{8 n^2} + %
\frac{l t }{2 n} \r) dt
\label{eq:EULER-MACLAURIN}
\end{equation}
 If we denote by $I_n$ the integral, we have after substituting $t =
2 n^{2/3}e^z$:
\begin{equation}
I_n \sim 2^{a+1} n^{\frac{2(a+1)}{3}} \int_{-\infty}^{+\infty} %
\exp(h(z)) dz 
\end{equation}
where 
\begin{equation}
h(z) = - \frac{1}{3} e^{3z} + \frac{l}{2 n^{2/3}} e^{2z} + %
\frac{l}{n^{1/3}} e^z + (a+1)z \, .
\label{eq:H}
\end{equation}
We have
\begin{equation}
h^{'}(z) = - e^{3z}  + \frac{l}{n^{2/3}} e^{2z} + %
\frac{l}{n^{1/3}} e^z + (a+1)
\label{eq:H1}
\end{equation}
and
\begin{equation}
h^{''}(z) = - 3 e^{3z} + \frac{2 l}{n^{2/3}} e^{2z} + %
\frac{l}{n^{1/3}} e^z \, .
\label{eq:H2}
\end{equation}
Let $z_0$ be the solution of $h^{'}(z) = 0$. $z_0$ is located near
$\frac{1}{3} \ln (a+1)$ because $a=(3l+1)/2$ is large.
Note that $z_0$ can be obtained solving
the cubic equation $h^{'}(z)=0$. Straightforward calculations leads to the
following estimate 
\begin{equation}
z_0 = \frac{1}{3} \ln \l(\frac{3}{2} (l+1) \r) %
+ O\l(\frac{l^{1/3}}{n^{1/3}} \r)  %
    =  \frac{1}{3} \ln \l( a+1 \r) %
+ O\l(\frac{l^{1/3}}{n^{1/3}} \r) \,
\label{eq:Z_0}
\end{equation}
and 
\begin{equation}
h(z_0) = \frac{a+1}{3}\ln{(a+1)} -  \frac{a+1}{3} %
+ O(\frac{l^{4/3}}{n^{1/3}}) \, .
\label{eq:hZ_0}
\end{equation}
We have also $h^{''}(z_0) = - \l( 3(a+1) + 2 l / n^{1/3} e^{z_0} + 
l/n^{2/3} \r) $  and more generally 
\begin{equation}
h^{(m)}(z_0) = -3^{m-1}(a+1) + A_m \frac{l}{n^{1/3}} e^{z_0} + %
B_m \frac{l}{n^{2/3}} e^{2 z_0} \, .
\label{eq:HMZ0}
\end{equation}
Thus,
\begin{equation}
\int_{-\infty}^{+\infty} e^{h(z)}dz = e^{h(z_0)} %
\int_{-\infty}^{+\infty} \exp \l( h^{''}(z_0) \frac{(z-z_0)^2}{2} %
 + P(z-z_0) \r) dz
\label{INTEGRALS}
\end{equation}
where $P$ is a power series of the form $P(x) = (a+1) \sum_{i \geq 3} p_i
x^i$ and $h^{''}(z_0) <0$. At this stage, one can consider 
$\exp \l( h^{''}(z_0) \frac{(z-z_0)^2}{2} \r)$ as the main
factor of the integrand. We refer here to the book of \auteur{De Bruijn}
 \cite[\S 4.4 and \S 6.8]{De Bruijn} for more discussions about asymptotic 
estimates on integrals of the form given by (\ref{INTEGRALS}) and we infer that
\begin{equation}
\int_{-\infty}^{+\infty} e^{h(z)} dz \sim %
\sqrt{ - \frac{2 \pi}{h^{''}(z_0)}} %
\exp \l(h(z_0)\r) \, .
\label{eq:BEFORE-STIRLING}
\end{equation}
Using the Stirling formula for Gamma function, i.e., 
$\Gamma(t+1) \sim \sqrt{2 \pi t} \frac{t^t}{e^t}$
and the fact that $z_0$ is located near $\frac{1}{3} \ln (a+1)$,
$h(z_0) \sim \frac{(a+1)}{3} (\ln (a+1)-1) $ and $h^{''}(z_0) \sim -3(a+1)$,
we can see that (\ref{eq:BEFORE-STIRLING}) leads to (\ref{eq:LEMMA2}) which is
also the formula obtained by \auteur{Janson} in \cite{Ja2000}.
\ENDPROOF

To estimate $\alpha_l$, due to (\ref{eq:LEMMA12}), it is convenient to
compare the magnitudes of
\begin{equation} 
\sum_{k=1}^{n} \frac{k^{(3l+1)/2}}{n^{l+1}} %
\exp\l(- \frac{k^3}{24 n^2} + \frac{l k^2}{8 n^2} + \frac{l k }{2 n} \r)
\label{eq:A}
\end{equation}
and of
\begin{equation}
\sum_{k=1}^{n} \frac{k^4}{n^3} \, \frac{k^{(3l+1)/2}}{n^{l+1}} %
\exp\l(-\frac{k^3}{24 n^2} + \frac{l k^2}{8 n^2} %
+ \frac{l k }{2 n} \r)  \, . 
\label{eq:B}
\end{equation}
Also we need to compare (\ref{eq:A}) to the other ``error terms''
contained in the ``big-ohs'' of (\ref{eq:LEMMA12}).
Using the asymptotic value given by lemma \ref{LEMMA2}, we easily obtain
the estimates of the two quantities and we compute respectively 
$2^{(3l+3)/2} 3^{(l-1)/2} \Gamma((l+1)/2)$ for (\ref{eq:A}) 
and $2^{(3l+11)/2} 3^{l/2 + 5/6} \Gamma( (l+1)/2 + 4/3 )/ n^{1/3}$ for 
(\ref{eq:B}). Thus, the term ``$O(k^4/n^3)$'' in (\ref{eq:LEMMA12})
can be neglected if $l=o(n^{1/4})$ otherwise the quantity represented by
(\ref{eq:B}) is not \textit{small} compared to that represented by
(\ref{eq:A}). Similarly,  straightforward calculations using (\ref{LEMMA2})
show that the terms $k/n$, $1/k$, $l^3/k^2$,
$l^{1/2}/k^{1/2}$ and $(l+1)^{1/16}/k^{9/50}$ can also be
neglected.  Using Stirling formula for Gamma function,
 lemmas \ref{LEMMA1} and 
\ref{LEMMA2}, we have
\begin{equation}
\alpha_l \sim \frac{\rho_l}{2} 2^{\frac{3l+3}{2}}%
 3^{\frac{l-1}{2}} \Gamma(\frac{l+1}{2}) \, .
\label{eq:AVANT_THEOREM}
\end{equation}

After nice cancellations, it results that:

\begin{thm} \label{THEOREM_EXPECTED_CREATION}
In a randomly growing graph of $n$ vertices, if $l,n \rightarrow \infty$
but  \\ 
\noindent $ l=o(n^{1/4})$,
the expected number of transitions $l \rightarrow l+1$,
for all $l$-components, is $ \alpha_l \sim 1 $.
\ENDTHEOREM
\end{thm}
Note that in \cite[p 301--306, \S 16--18]{JKLP93}, the authors already
proved, by entirely different methods, that the most probable evolution
of a random graph, when regarding the excess of connected component,
is to pass directly from $1$-component to $2$-component, from
$2$-component to $3$-component, and so on.

Similarly, as an immediate consequence of calculations above and
\cite[Theorem 9]{Ja2000}, we have:

\begin{cor} \label{corollary}
As $n \rightarrow \infty$ and $l=o(n^{1/4})$, the expected number of vertices
that ever belong to an $l$-component is $\mathbb{E} V_l \sim (12l)^{1/3}
n^{2/3}$ and the expected order of the largest $l$-component that ever appears
is $\mathbb{E} \VMAX = O(l^{1/3} n^{2/3})$.
\ENDTHEOREM
\end{cor}
Note that these results answer the last remark in \cite{Ja2000}.

\section{Conclusion}
We briefly point out a remark concerning the restrictions on $l$ in 
constrained graphs problems, i.e., the creation and growth of
components with prefixed configurations. A possible way of 
investigations could be to search for similar 
results with those of \auteur{Bender-Canfield-McKay}
\cite{BCM90,BCM92} for supergraphs of a given graph $H$.

\bibliographystyle{plain}

\end{document}